# WiFi-TCN: Temporal Convolution for Human Interaction Recognition based on WiFi signal

Chih-Yang Lin[1], Senior Member, IEEE, Chia-Yu Lin[2], Yu-Tso Liu[2], and Timothy K. Shih[2], Senior Member IEEE

[1]Department of Mechanical Engineering, National Central University, Taoyuan City 32001, Taiwan
[2]Department of Computer Science and Information Engineering, National Central University, Taoyuan City 32001, Taiwan

Corresponding author: Timothy K. Shih (e-mail: tshih@g.ncu.edu.tw).

**ABSTRACT** The utilization of Wi-Fi based human activity recognition has gained considerable interest in recent times, primarily owing to its applications in various domains such as healthcare for monitoring breath and heart rate, security, elderly care. These Wi-Fi-based methods exhibit several advantages over conventional state-of-the-art techniques that rely on cameras and sensors, including lower costs and ease of deployment. However, a significant challenge associated with Wi-Fi-based HAR is the significant decline in performance when the scene or subject changes. To mitigate this issue, it is imperative to train the model using an extensive dataset. In recent studies, the utilization of CNN-based models or sequence-to-sequence models such as LSTM, GRU, or Transformer has become prevalent. While sequence-to-sequence models can be more precise, they are also more computationally intensive and require a larger amount of training data. To tackle these limitations, we propose a novel approach that leverages a temporal convolution network with augmentations and attention, referred to as TCN-AA. Our proposed method is computationally efficient and exhibits improved accuracy even when the data size is increased threefold through our augmentation techniques. Our experiments on a publicly available dataset indicate that our approach outperforms existing state-of-the-art methods, with a final accuracy of 99.42%.

## I. INTRODUCTION

This The recognition of human activity (HAR) in an indoor environment using Wi-Fi applications has emerged as a significant and widely researched field [2]. Literature review [3, 4, 5] indicates that camera-based and sensor-based HAR systems are widely utilized in addressing this issue, however, these approaches are often associated with high costs or the requirement of wearable devices, which can be uncomfortable and inconvenient for users [6]. Given that Wi-Fi has the potential to overcome these limitations, it is being considered as a promising alternative solution for HAR [1, 3, 7]. Although Wi-Fi has its advantages in indoor environments, it still faces certain challenges such as limited coverage and difficulties in handling multi-subject scenarios.

In the realm of Human Activity Recognition (HAR), there are three distinct types of Wi-Fi signals that are employed, namely hardware-based radio signals, Received Signal Strength Indicator (RSSI), and Channel State Information (CSI). The utilization of RSSI signals has garnered widespread attention in a multitude of sensing applications, including indoor positioning and tracking [8, 9]. However, due to the inherently coarse nature of RSSI signals, it poses a challenge to attain high levels of precision in fine-grained HAR. For instance, certain interactions as described in [1] that entail fine-grained activities such as "kicking with the left leg", "kicking with the right leg", "pointing with the left hand", and "pointing with the right hand" are particularly difficult to accurately determine due to the limitations posed by RSSI signals.

The Channel State Information (CSI) provides detailed information at the physical layer that is invaluable for precise activity recognition. The utilization of techniques such as Multi-Input Multiple-Output (MIMO) and Orthogonal Frequency Division Multiplexing (OFDM) [10] allows for the acquisition of additional phase and amplitude information of each sub-carrier from the Wi-Fi signal that is spread between the transmitting and receiving antennas and operates at a specific carrier frequency. As a result of limb movement, variations in wireless signals can be observed. Early research has established the superiority of CSI compared to the traditional Received Signal Strength Indicator (RSSI) [11]. Despite the ability of CSI to capture the aggregate impact of multipath interference caused by the human body, variations in the signals may still persist, even when the same subjects are engaged in the same activity. To mitigate this issue, it is crucial to gather a substantial amount of data to prevent the model from unduly focusing on differences between subjects rather than the activity itself. To this end, some researchers have utilized data augmentation to increase the diversity of the dataset [12, 13], though the outcomes remain unsatisfactory.

Our main contributions are as follows:
1) We propose a novel model known as Temporal Convolution Network with Augmentations and Attention (TCN-AA). Our approach diverges from conventional Convolutional Neural Networks (CNNs) and sequence-to-sequence models such as Long-Short Term Memory (LSTM) [32] and Gated Recurrent Unit (GRU), instead relying on a temporal convolution network. By utilizing this architecture, we are able to reap the benefits of convolution networks, such as the capacity to process sequential inputs, as well as significantly reduced training duration.
2) We incorporate data augmentation techniques on Wi-Fi signals to expand the available data, and implement an attention mechanism to enhance the model's focus on the activity and expedite convergence.
3) The empirical results of our proposed experiments demonstrate that TCN-AA outperforms existing state-of-the-art methods and achieves the highest accuracy.

The remainder of this paper is organized as follows. Section II presents a comprehensive overview of the previous studies on WiFi-based Human Activity Recognition (HAR). In Section III, We provide a detailed description of our environment setup and system design, including our proposed model, Temporal Convolution Network with Augmentations and Attention, and an attention mechanism to enhance the model's focus on the activity and expedite convergence. In Section IV, conduct a comprehensive evaluation of our model and method, including the TCN-AA model, attention mechanism of expedite convergence, and compare it to existing approaches. Finally, the conclusion and potential directions for future research are presented in Section V.

## II. RELATED WORK

A comprehensive overview of recent studies pertaining to Wi-Fi-based Human Activity Recognition (HAR) will be presented. The available approaches for Wi-Fi-based HAR can be mainly divided into two subgroups, based on their methodologies: Convolution Neural Networks (CNNs), and Sequence-to-Sequence (Seq2Seq) models.

### A. CNN-BASED APPROACHES

The majority of early studies on Wi-Fi-based human activity recognition (HAR) employ Convolutional Neural Network (CNN) based models as a starting point. This is due to the characteristic pattern of activity present in the Channel State Information (CSI) signals, which make CNNs an appropriate choice for recognition. For instance, Wang et al. [14] utilized a modified U-Net of deep convolutional neural network and annotations on 2D images to perform body segmentation and pose estimation from CSI signals. Their model was evaluated on more than 105 frames across 16 indoor scenes and achieved scores of 0.91 for AP@50, 0.65 for mIoU, and an average mAP score of 0.38 over AP@50-AP@95. While their performance is not optimal, their study demonstrated the potential to reconstruct fine-grained 2D spatial information of human bodies from CSI signals.

Kabir et al. [15] developed the CSI-based Inception Attention Network (CSI-IANet) incorporating CNNs and spatial-attention and evaluated it using a dataset of Wi-Fi-based human-to-human interactions (HHI), which is the same dataset used in this paper [1]. The HHI dataset includes 12 different human-to-human interactions performed by two subjects and will be described in detail in section III. The CSI-IANet achieved an average accuracy of 91.3%, making it the first CNN-based model to surpass 90% accuracy on this dataset. The authors applied a Butterworth low-pass filter to denoise the CSI signal, employed an inception module to provide the model with varying receptive fields, and utilized spatial attention to achieve such high accuracy.

Subsequently, Shafiqul et al. [2] proposed the HHI-AttentionNet, which employed a depth-wise CNN and a customized attention mechanism, achieving 95.47% accuracy on the same dataset [1]. The authors applied a Butterworth low-pass filter to remove significant peaks, such as outliers and high-frequency noise, and a Gaussian smoothing function to remove short peaks. Both CSI-IANet and HHI-AttentionNet utilized segmentation to split the data into smaller windows. However, this approach can compromise the integrity of the data as it requires fixing the length of each sample, and the computing power required increases as the length of each sample decreases. In contrast, this paper fixes the length of each sample by chopping some parts of the steady state and applying a one-dimensional Discrete Wavelet Transform (DWT) to downsample the data, which will be described in detail in Section III. Despite the high accuracy of 95.47% achieved by HHI-AttentionNet, there is still room for improvement. Among the studies utilizing the same dataset [1], the 95.47% accuracy is the highest achieved by a CNN-based model and it appears to be the limit for CNN-based models.

### B. SEQ2SEQ MODEL APPROACHES

The limitations of Convolutional Neural Network (CNN) models have led researchers to investigate alternative approaches for processing time-sequence data. One such approach is the use of Sequence-to-Sequence (Seq2Seq) models. In recent years, there has been an increasing trend in the use of Seq2Seq models for Wi-Fi-based Human Activity Recognition (HAR).

An attention-based bi-directional long short-term memory (ABLSTM) network approach has been proposed in [16] for the implementation of a passive human activity recognition system. The system utilizes channel state information obtained from Wi-Fi signals to recognize six common daily activities including falling, walking, running, standing up, sitting down, and lying down. To achieve this, the authors first employ a bi-directional LSTM to encode the CSI signal, and then use attention mechanism on the LSTM output to make the final prediction. The proposed approach demonstrates remarkable performance with an average accuracy of 96.7% and 97.3% on two different environments, namely the Activity Room and the Meeting Room. However, it is worth noting that the high accuracy achieved in this study is based on data collected from a single individual performing coarse-grained activities, which may have made the recognition task easier for the model.

A different Seq2Seq approach is Two-Stream Convolution Augmented Human Activity Transformer (THAT) [17] that utilizes CSI signals to recognize seven daily activities [16], including an additional activity, "picking up." The authors propose a network architecture that combines the strengths of convolutional neural networks and multi-head self-attention transformers to achieve high recognition accuracy. The architecture utilizes two distinct input streams, a temporal stream and a channel stream, which are fed into separate CNN-based transformers. This design allows the model to generate distinct features through convolution in the temporal and channel domains, ultimately leading to improved performance. The results of the study show that the use of two input streams significantly improves recognition accuracy, with an average accuracy of 98.6% across four different scenes. Although this approach is faster than state-of-the-art Seq2Seq-based methods, it still requires a substantial amount of time for training and only uses data from a single subject.

Another example of a device-free deep learning (DL) model, H2HI-Net [18], presents an illustration of a multimodal approach to tackle the human-to-human interaction problem. The authors propose a combined model that encompasses a residual neural network and a bi-directional gated recurrent unit (Bi-GRU), which is evaluated using the HHI dataset [1]. Preprocessing of the CSI signal involves denoising through the application of the Butterworth filter and dimensionality reduction via principal component analysis. The residual neural network and Bi-GRU are then utilized to encode the spatial and temporal embeddings, respectively. The interactions are finally predicted through the concatenation of these two embeddings with a two-layer Dense Net, resulting in an average accuracy of 96.39%. While this performance is satisfactory, the authors still employ the Seq2Seq approach, which results in increased computational demands. Conversely, this paper utilizes a temporal convolution network (TCN) with an attention mechanism and data augmentation, resulting in improved accuracy exceeding 96.39% while ensuring low computational requirements.

## III. MATH

This section provides an overview of the dataset, the proposed methods, and the corresponding model. Initially, a description of the dataset is presented, which encompasses details on the interactions during data collection, the environment, and the environmental settings. Subsequently, the preprocessing and augmentation techniques employed in the proposed methods are provided. Finally, the section concludes with a discussion of the temporal convolution network model and the hyperparameters that resulted in optimal performance. The methodology of the proposed approach is presented in a flowchart in Fig. 1.

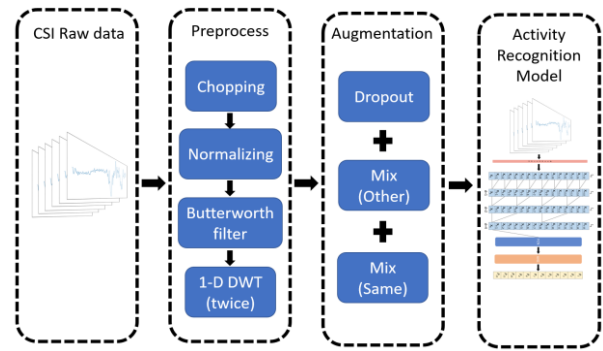

**FIGURE 1. The flowchart of our methods.**

### A. CSI DATASET OF HUMAN-TO-HUMAN INTERACTION

In this paper, the proposed method is validated using the publicly available CSI dataset [1]. This dataset comprises 64 subjects consisting of 40 distinct pairs, who were required to perform a total of 12 interaction activities (i.e., $N_c = 12$), including approaching, departing, handshaking, high five, hugging, kicking with the left leg, kicking with the right leg, pointing with the left hand, pointing with the right hand, punching with the left hand, punching with the right hand, and pushing. The subjects performed 10 trials for each activity, with each trial lasting between 5 to 6 seconds, including a 2-second steady state. The packet length ($N_p$) varied between [1040, 2249].

Data collection was performed using a commercial off-the-shelf access point, such as the Sagemcom 2704 Wi-Fi router, as the transmitter and a desktop computer equipped with an Intel 5300 NIC as the receiver. The MIMO Wi-Fi streams consisted of 2 internal antennas at the transmitter (Nt = 2) and 3 external antennas at the NIC (Nr = 3). Data was collected using a bandwidth of 20 MHz and a frequency of 2.4 GHz, and the CSI was captured using a publicly available CSI tool [22] with 30 subcarriers (i.e., Ns=30).

The Intel 5300 NIC chipset utilizes the 802.11n channel state information to present channel matrices, each having a signed 8-bit resolution for the real and imaginary parts. In essence, each transmitter-receiver pair derived from the CSI signals can be expressed as H(Ti, Rj), where $h_y^x$ represents the complex number at the xth packet and yth subcarrier, with $x \in [1, \ldots, Np]$ and $y \in [1, \ldots, Ns]$. The notation (Ti, Rj) signifies that the signal is transmitted from the ith antenna at the transmitter and received by the jth antenna at the receiver. The CSI signals can be viewed as a vector as:

$$H(T_i, R_j) = \begin{bmatrix} h_1^1 & \cdots & & h_N^1 \\ & \ddots & & \\ \vdots & & h_y^x & \vdots \\ & & & \ddots \\ h_1^{N_p} & \cdots & & h_{N_s}^{N_p} \end{bmatrix}$$

*B. PREPROCESSING*

The task of directly recognizing CSI signals from their values proves to be a challenge due to the inherent noise present in CSI signals. Fluctuations in CSI values are a common occurrence, even in environments without any human presence, due to the presence of internal and external sources of interference. The internal interferences include transmission rate adaptations and variations in power, while external interferences are comprised of background radio noise and interference caused by moving objects. As such, it is imperative that the CSI values undergo preprocessing prior to further analysis.

During the data collection process, the transmitter transmits data to the receiver at a specified frequency, which can result in fluctuations in the actual sampling rate due to internal and external sources of interference. At times, the overhead demands of the device may be excessive, leading to a significant drop in the sampling rate for a short duration. Data collected in this manner is often considered flawed due to a lower number of total packets in comparison to other sets of data. Our dataset encompasses packets of data ranging in size from 1040 to 2249. In order to discard data with fewer than 1500 packets, we have established a threshold of 1500 (i.e., Np=1500). For data with a packet count exceeding 1500, we have implemented a process of chopping the steady state in order to standardize the packet count at 1500. The final input data is represented in the form of (Nt × Nr, Np, Ns).

Despite the consistency in the activity being performed by the same subject-pair, variations in the CSI value can occur due to external interference. To mitigate this issue, normalization is imperative. The normalization process involves transforming the data to a range of [-1,1] for each Transmitter-Receiver pair. Subsequently, a Butterworth filter, a low-pass filter that effectively eliminates most high-frequency noise, is applied to each TR-pair. Finally, to achieve down sampling, a one-dimensional Discrete Wavelet Transform is applied twice, reducing Np from 1500 to 375. In the process of utilizing one-dimensional Discrete Wavelet Transform (1D-DWT), the original signal undergoes decomposition into two sets of coefficients: the approximation coefficients and the detailed coefficients. The approximation coefficients depict the low-frequency information, while the detailed coefficients correspond to high-frequency information. Given that the most significant interactions occur in the low-frequency domain, we choose to retain the approximation coefficients and discard the detailed coefficients.

The efficacy of our methodology is illustrated in Fig. 2, where we present the amplitude data of three separate interactions. The application of a Butterworth filter led to a substantial reduction in high-frequency noise. Furthermore, repeating the 1D-DWT process resulted in unchanged feature patterns and a decrease in both data length and training time. This is of significant value for researchers without access to advanced GPUs. Additionally, 1D-DWT plays a crucial role in removing high-frequency noise, as demonstrated in our experimental section.

*C. AUGMENTATION*

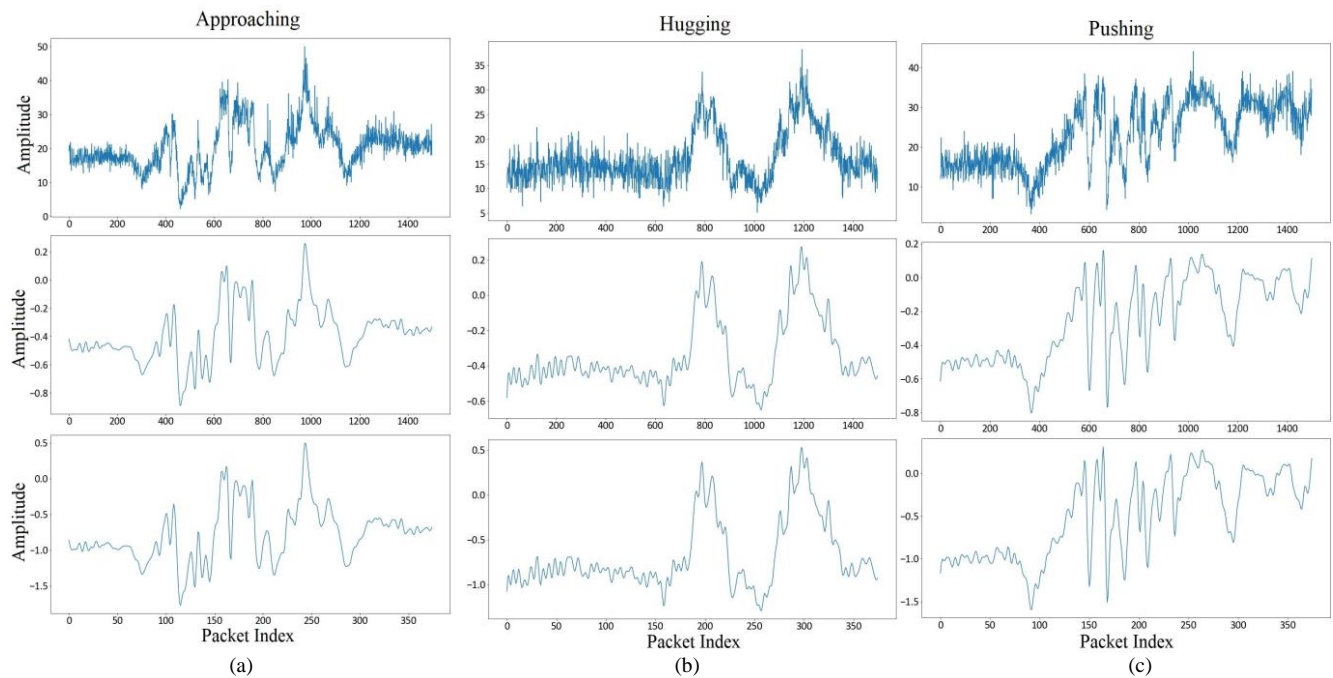

FIGURE 2. The flowchart of our methods. (a) is the amplitude data of interaction "Approaching", (b) is the amplitude data of "Hugging", and (c) is the amplitude data of "Pushing". For the figures in first row, they represent the raw data. For the figures in middle row, they represent the data after a Butterworth lowpass filter and normalization. For the figures in third row, they represent the data after applying 1D DWT twice. As you can see, the length of packets in third row is 375.

In contrast to the abundance of datasets available for image recognition, the availability of datasets pertaining to Wi-Fi is relatively scarce. This disparity is demonstrated by the fact that datasets for image recognition, such as ImageNet and COCO [33], contain an order of magnitude greater number of samples compared to Wi-Fi-based datasets. Thus, the implementation of data augmentation techniques for Wi-Fi-based datasets is of utmost importance. Our research endeavors addressed this issue by employing and comparing the efficacy of three distinct augmentation techniques. The methods employed are presented as follows:

1) DROPOUT
In our proposed method, a portion of the CSI values was randomly set to zero with a probability λ, which was selected randomly from the range of (0, 0.07). Based on the established success of dropout in deep learning, it is reasonable to assume that its application on raw data as an augmentation technique will prove to be effective. The utilization of dropout on raw data is thus proposed as a promising method of augmentation in this context.

2) MIX SAMPLES WITH DIFFERENT LABELS
The presence of external interferences, such as background radio noise and disturbances from moving objects, has been previously described. The intensity of background radio noise varies dynamically over time. To mitigate this issue, a technique of mixing samples with diverse labels has been proposed as a means of enhancing the robustness of the model. The following equation illustrates the mixing procedure:

$$D = A \cdot (1 - \varepsilon_1) + B \cdot \varepsilon_2 + C \cdot \varepsilon_3. \quad (1)$$

where the new sample, D, is obtained as a result of mixing sample A, B, and C. The label of D is inherited from sample A, while the labels of samples B and C differ from that of sample A. In the course of our experiments, the value of $\varepsilon_k$ was randomly selected from the range of (0, 0.05).

3) MIX SAMPLES WITH SAME LABELS
This method utilizes the same mixing equation as presented in (1). The primary distinction is the mixing of samples that possess equivalent labels. This approach is motivated by the observation that moving object disturbance is a result of variations in subjects' heights and body shapes, even when they are performing the same activity. Given that moving object disturbance is expected to be more pronounced than background radio noise, it was deemed necessary to mix samples with the same label to enhance the robustness of the model.

*D. MODEL*
The CSI signal exhibits a strong correlation with time-sequence data, making it well-suited for the application of time-sequence deep learning models such as Long Short-Term Memory (LSTM). The LSTM model operates through the use of three gates: the input gate, the forget gate, and the output gate. These gates enable the model to retain past memories; however, they also require substantial computational resources, resulting in longer training times. In contrast, the Temporal Convolution Network (TCN) [19, 20] presents a more suitable solution for our needs. The TCN utilizes one-dimensional causal convolution and dilated convolution in the temporal domain, making it a more efficient model compared to LSTM.

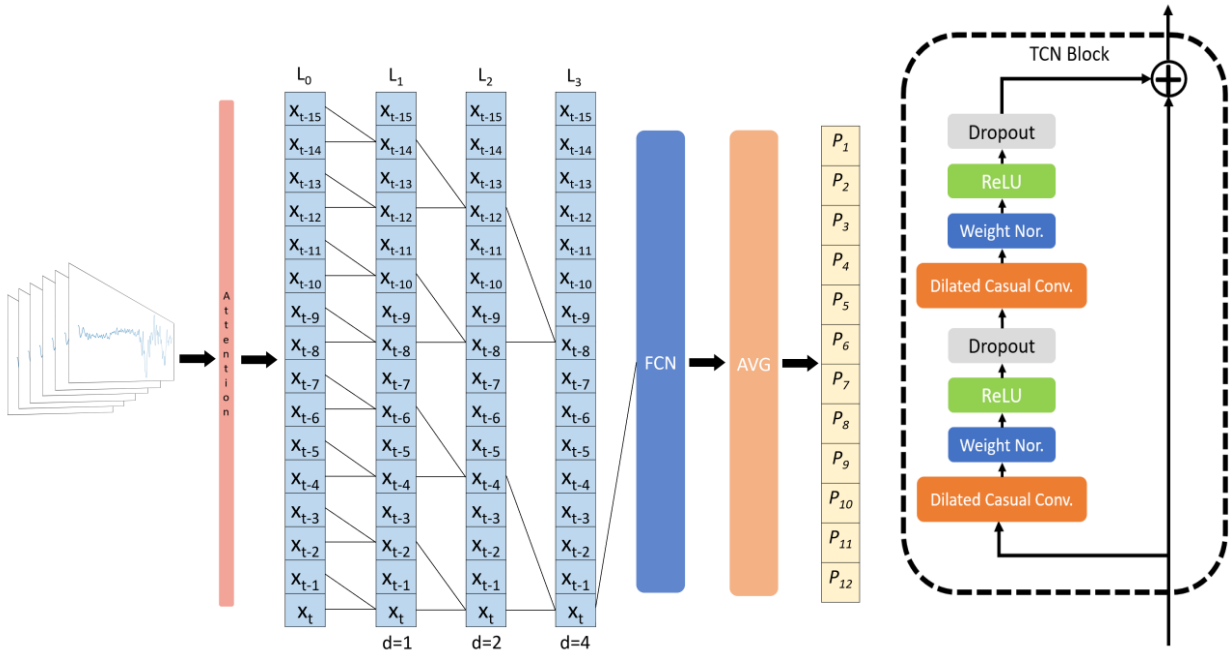

FIGURE 3. The architectural overall of our system in this paper. (a) An overview of the TCN model which give an example of kernel size = 2. The $L_0$ represents the input after attention. The dilation size of $L_1$, $L_2$, and $L_3$ is 1,2, and 4, respectively. The output, $P_1$ to $P_{12}$, is the probabilities of 12 classes. (b) A detail of each TCN layer.

We identify three main advantages of the TCN model. Firstly, the convolution operation of TCN is highly effective in extracting features from CSI signals, as evidenced by previous research in the field. Secondly, the convolutional operation of TCN requires significantly less computational resources, resulting in shorter training times compared to LSTM. Lastly, the causal and dilated convolution of TCN enables the model to gather information from past data. In the following sections, we will provide a comprehensive overview of our proposed model and the TCN, as depicted in Fig. 3.

### 1) CAUSAL CONVOLUTIONS

It is a commonly accepted practice in sequence-to-sequence models to maintain the same length of output as input, as well as to preserve past information. The Temporal Convolution Network (TCN) accomplishes this goal through the use of causal convolutions. To implement the TCN model, a full 1-dimensional convolutional network architecture is employed, with a kernel size of k and a padding size of " k - 1". The padding is accomplished using zeros, ensuring that each layer's output maintains the same length as the input, as demonstrated in Fig. 4. This is due to the fundamental principle of TCN, which prevents future information from leaking into the past. As a result, the output at time t is always convolved only with information from time t and earlier.

### 2) DILATED CONVOLUTIONS

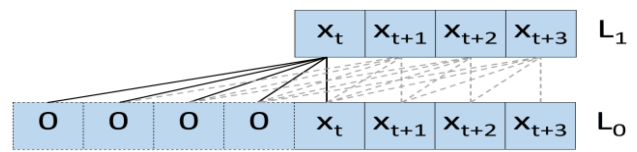

FIGURE 4. An example of 1D causal convolution while the kernel size is 5, the padding size is 4, and padding with 0.

The utilization of sequence-to-sequence models presents a challenge in terms of memory capacity. While an increase in the model's ability to retain historical information results in improved performance, it also leads to a larger memory footprint. To address this issue, two possible solutions have been proposed. The first approach entails the utilization of large kernel sizes in causal convolutions; however, this option appears to be impractical. The alternative solution, as introduced by Yu and Koltun [24], involves the use of dilated convolutions, which have since become a widely utilized technique. The dilated mechanism enables the TCN to possess an exponentially expanding receptive field, as demonstrated in Fig. 5. In our experiments, the dilated size was initially set at 1 and grew exponentially with a base of 2. For instance, in our model, there are three layers of TCN, each with a dilated size of 1, 2, and 4 respectively. Furthermore, we increased the kernel size (k=15) in order to maintain a larger receptive field. Although training with a larger kernel size may take longer, it remains a more cost-effective solution compared to the use of LSTM networks.

The comparison between the performance of LSTM and TCN will be discussed in Section IV.

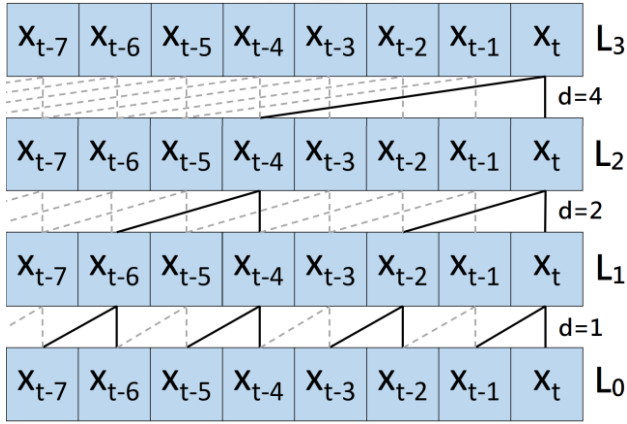

**FIGURE 5.** An example of dilated convolution while the kernel size is 2 and the delated size increases from 1 to 4 exponentially.

### 3) TEMPORAL ATTENTION MECHANISM

The widespread success of transformers in recent years is largely attributed to their utilization of the attention mechanism. The implementation of an appropriate attention mechanism can result in both an improvement in accuracy and an acceleration of convergence speed during the training process. Our paper adopts a mechanism of attention that is primarily inspired by the works of Hao et al. [20] and Vaswani et al. [25]. The input (H) is first linearly transformed into three distinct vectors, named query (Q), key (K), and value (V), respectively, as expressed by the following formulas:

$$Q = W_Q \cdot H. \quad (2)$$
$$K = W_K \cdot H. \quad (3)$$
$$V = W_V \cdot H. \quad (4)$$
$$\text{Attention}(Q, K, V) = \text{softmax}\left(\frac{QK^T}{\sqrt{d_K}}\right) V. \quad (5)$$
$$H' = H \cdot \text{Attention}(Q, K, V). \quad (6)$$

The weight of the linear transformation, represented by WQ, WK, and WV, is learned during the training phase. The outcome of the softmax operation performed on the product of the query (Q) and key (K) serves as a weighted factor which is divided by the square root of dK, the dimension of key (K), and then multiplied with the value (V). This attention mechanism allows the model to dynamically concentrate on the most pertinent features of the input during the prediction process. Upon computing the attention score, it is multiplied with the input (H) to produce a new input (H'). Subsequently, some modifications are made to Eq. 5 in accordance with the principles of causal convolutions. As previously mentioned, the fundamental principle of causal convolutions is that future information must not penetrate into the past. Thus, there is no requirement for calculating the correlation between the past and the future. As a result, Eq. 5 is revised as follows:

$$\text{Attention}(Q, K, V) = \text{softmax}\left(L\left(\frac{QK^T}{\sqrt{d_K}}\right)\right) V. \quad (7)$$

where the function L represents the lower triangular function. Upon performing the multiplication of the query and keys, and division by the square root of dK, the values above the main diagonal in the resulting matrix are set to zero, as demonstrated in Fig. 6. In consideration of time and performance constraints, we have limited the incorporation of attention to the first layer, as depicted in Fig. 3. The impact of adding attention at various layers will be assessed in Section IV for comparison purposes.

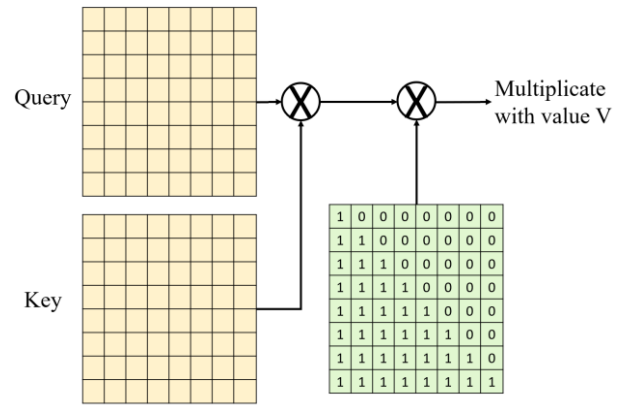

**FIGURE 6.** An example of the lower triangular function.

### 4) FCN AND AVERAGE POOLING

Our proposed model consists of three hierarchical layers of TCN blocks. The number of filters utilized in each layer has been set to a uniform value of 50 (i.e., $N_f^{L_1} = N_f^{L_2} = N_f^{L_3} = 50$). As a result, the output of TCN is a three-dimensional vector which is in the shape of (Nt × Nr, Np, $N_f^{L_3}$). The first dimension represents the TR-pair, the second dimension represents the temporal dimension, and the third dimension represents the frequency domain. Only the last term of the second dimension is subsequently fed into a fully connected network (FCN) [23], which takes the shape of (Nt × Nr, 1, $N_f^{L_3}$). The output of the FCN is then represented as (Nt × Nr, 1, Nc). Considering the existence of Nt × Nr TR-pairs, each with a distinct signal, we interpret the output of the FCN as Nt×Nr individual probability distributions, each pertaining to one of the 12 classes, based on the respective TR-pair. In this scenario, we apply an average pooling on the Nt × Nr probability distributions to yield the final output of the model, which is of the shape (1, 1, Nc).

### E. VARIABLE AND HYPERPARAMETERS

Here we provide a summary of the symbols and variables utilized, along with the experimentally determined hyperparameters that were chosen to achieve optimal performance.

| Variable | Meaning |
|---|---|
| $N_c$ | The number of classes (i.e., 12). |
| $N_p$ | The number of packets. |
| $N_t$ | The number of antennas at the transmitter (i.e., |

|  | 2). |
|---|---|
| $N_r$ | The number of antennas at the receiver (i.e., 3). |
| $N_s$ | The number of subcarriers (i.e., 30). |
| $h_y^x$ | A complex number corresponding to the $x^{\text{th}}$ packet and the $y^{\text{th}}$ subcarrier for a particular TR-pair. |
| $H(T_i, R_j)$ | A CSI value transmitted by the $i^{\text{th}}$ transmitter and received by the $j^{\text{th}}$ receiver. |
| $\lambda$ | The dropout rate in data augmentation. |
| $\varepsilon_k$ | The mixing rate in data augmentation where $k \in [1, 2, 3]$ in our method. |
| $N_f^{Lm}$ | The number of filters in the $m^{\text{th}}$ layer of TCN where $m \in [1, 2, 3]$ in our model. |

| Hyperparameter | Value |
|---|---|
| Number of TCN layer | 3 |
| The number of filters at each layer | [50, 50, 50] |
| Kernel size ($k$) | 15 |
| Attention Mechanism | Add at 1st layer only |
| Dropout rate | 0.5 |
| Batch size | 32 |
| Optimizer | AdamW (Adam + weight decay) |
| The rate of exponential decay of learning rate | 0.988/per epoch |
| Epoch | 200 |
| K-fold | 10 folds |

## IV. EXPERIMENTS

In this section, a comprehensive experimental analysis will be performed to evaluate various aspects of our proposed method. The first comparison will be against the performance of a Long-Short Term Memory (LSTM) model that has been trained by ourself. Subsequently, a comparison will be made with the results of other models that have been implemented on the HHI dataset. As stated in Section III, the dataset consists of 12 unique interactions, each performed by 40 subject pairs for 10 trials, yielding a total of 400 samples per interaction (class). Finally, the performance of the model will be further scrutinized through a series of ablation studies that focus on factors such as augmentation, kernel size, dropout, and attention.

### A. COMPARE TO LSTM

Initially, we trained the HHI dataset using a bidirectional LSTM network. The architecture of the LSTM network is depicted in Fig. 7, consisting of two layers with 180 neurons per layer. The LSTM was followed by a fully connected neural network [23] that outputted the likelihood of the 12 classes based on the features extracted by the LSTM. As demonstrated in Fig. 8, the accuracy of the LSTM model was approximately 81%, and its training time was three times longer compared to our proposed TCN model.

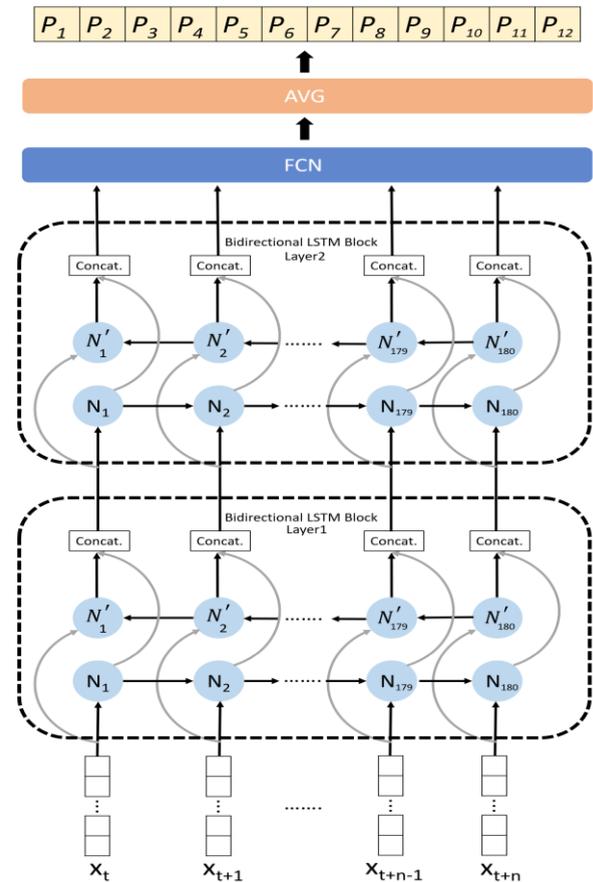

FIGURE 7. The architecture of LSTM model we used to compare.

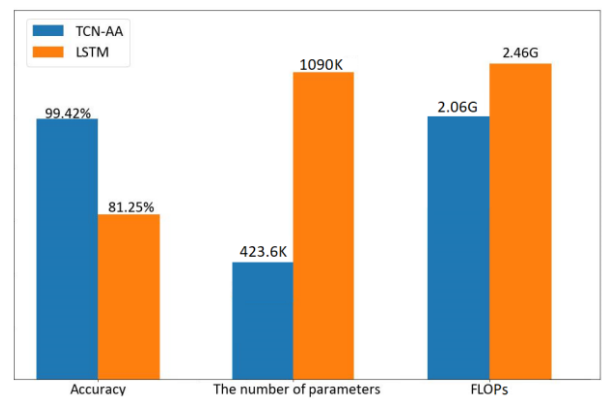

FIGURE 8. The comparison between our TCN-AA model and LSTM.

### B. COMPARE TO STATE-OF-THE-ART-METHODS

In this study, we will compare the performance of our TCN model with that of other methods that were applied to the same HHI dataset. These methods can be broadly categorized into two groups: traditional methods and deep learning-based methods. A customized support vector machine (SVM) proposed by [26] utilized principal component analysis (PCA) as preprocessing and the minimum redundancy maximum relevance (MRMR) algorithm to select a subset of the extracted features. Then, they applied SVM for classification, achieving an accuracy of 86.21%.

With respect to deep learning-based methods, Kabir et al. [15] introduced CSI-IANet, which utilized a Butterworth low-pass filter to denoise the CSI signal, employed three layers of CNN which is an inception module providing the model with varying receptive fields, and spatial-attention to first achieved an accuracy over 90% (91.3%). Kabir and Shin [27] presented DCNN, which employed only three layers of CNNs and achieved an accuracy of 88.66%. Shafiqul et al. [2] proposed HHI-AttentionNet, which utilized depth-wise CNNs and a customized attention mechanism, achieving an accuracy of 95.47%. Hao et al. [28] proposed a new Gabor residual block for CNN, combined with temporal attention and frequency attention, resulting in an accuracy of 86%. Alazrai et al. [29] proposed E2EDLF, which employed two layers of CNNs and achieved an accuracy of 86.3%.

Two additional methods based on sequence-to-sequence models were described. Shams et al. [30] presented Attention- BiGRUs, which used bidirectional gated recurrent units (Bi-GRUs) and multi-head attention, achieving an accuracy of 87%. Abdel-Basset et al. [18] proposed H2HI-Net, which utilized three layers of 1D CNNs with a residual network and BiGRUs, achieving an accuracy of 96.39%. As shown in Fig. 9, our TCN model achieved the best performance with an accuracy of 99.42%, outperforming H2HI-Net by 3%.

The superiority of deep-learning-based methods over SVM methods is evident in our proposed approach. The proposed augmentation methods and the use of the TCN model contribute to this advantage. The application of a causal convolution and a dilated convolution enables more efficient feature extraction in the temporal domain. Even without the proposed augmentation methods, our model demonstrates a high level of accuracy, registering 89.67%, which surpasses the performance of a significant number of other models. In comparison to sequence-to-sequence models, our model offers a higher level of efficiency while maintaining a smaller number of parameters, leading to reduced power consumption.

### C. ABLATION STUDY
#### 1) AUGMENTATION
As outlined in Section III, the proposed augmentation methods have resulted in a doubling of the amount of data. The raw data was utilized to evaluate performance, with the hyperparameters remaining the same, except for the kernel size, dropout rate, and attention. In this particular case, no attention networks were applied to the augmentation methods, and their performance was the sole focus of evaluation. The model was configured with a kernel size of 2 and a dropout rate of 0.2.

A comparison of three augmentation methods is presented, namely, dropout, mixing with the same label (i.e., Mix (Same)), and mixing with different labels (i.e., Mix (Other)). In the experiment, there were 400 samples in raw data for each class, and after each augmentation method, an additional 400 samples were generated for each class. As depicted in Table I, the highest accuracy achieved using raw data alone was only 57%. However, the application of the different augmentation methods resulted in a significant increase in accuracy, with ranges between 88.54% and 90.18%. This demonstrates that the use of the augmentation methods can result in an improvement of around 30% in accuracy over the use of raw data alone.

### Table I

| Augmentation Method | Accuracy(%) Training | Accuracy(%) Validation | Loss Training | Loss Validation |
|---|---|---|---|---|
| A | 76.96 | 56.81 | 0.93 | 1.31 |
| A+B | 88.54 | 85.21 | 0.67 | 0.75 |
| A+C | **90.18** | **89.69** | 0.65 | **0.69** |
| A+D | 90.09 | 88.74 | **0.64** | 0.73 |
| (A+B)/2 | 82.20 | 69.15 | 0.85 | 1.18 |
| (A+C)/2 | **82.47** | 69.57 | 0.84 | **1.16** |
| (A+D)/2 | 82.44 | **70.00** | **0.83** | 1.12 |

**Table I**. The comparison while doubling the amount of the data and fixed amount of the data. *A* represents the raw data. *B*, *C*, and *D* represent the method of dropout, mix(other), and mix(same), respectively. (A+B)/2 represents selecting a half of the amount of the data randomly.

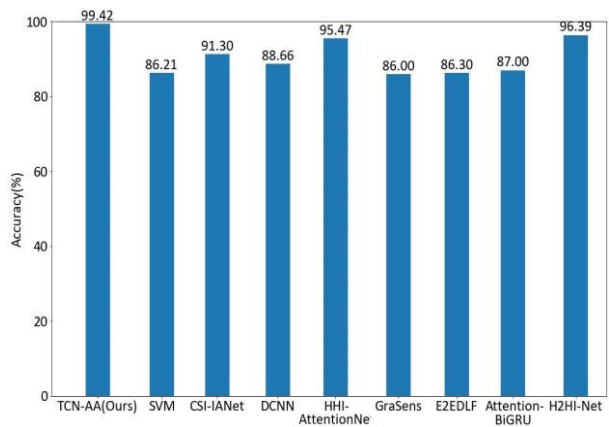

**FIGURE 9.** The comparison over other models.

In order to evaluate the quality of the augmented data, experiments were conducted with a fixed amount of data. As previously mentioned, each augmentation method generated an additional 400 samples per class, for a total of 800 samples per class. Then, 400 samples were randomly selected out of the 800, and their results were compared with the results of using only raw data alone. As shown in Table I, the average accuracy of each method was found to be 13% higher than using only the raw data alone. These results indicate that the augmented data not only provides greater diversity but also improved quality.

Furthermore, we found that the impact of the amount of "Approaching" and "Departing" classes (class 1 and 2) on accuracy are insignificant, as shown in Fig. 10. While the number of samples in class 1 and 2 are 800, the highest accuracy is 99.17%. The number of samples in class 1 and 2 are 1600, the highest accuracy is 99.42%.

of dropout. During the training phase, only 50% of the features are utilized for classification, while in the validation phase, 100% of the features are used.

Table II

| Kernel Size | Accuracy(%) Training | Accuracy(%) Validation | Loss Training | Loss Validation |
|---|---|---|---|---|
| 2 | 62.55% | 54.68% | 1.17 | 1.28 |
| 3 | 73.87% | 65.32% | 0.95 | 1.08 |
| 7 | 92.75% | 85.96% | 0.40 | 0.53 |
| 11 | 96.88% | 89.36% | 0.26 | 0.42 |
| 15 | 98.16% | 90.21% | **0.21** | **0.39** |
| 19 | 98.28% | **90.64%** | **0.21** | 0.40 |
| 23 | 98.42% | 90.00% | 0.22 | 0.46 |
| 27 | 97.85% | 89.57% | 0.23 | 0.45 |
| 31 | **98.44%** | 90.43% | 0.22 | 0.46 |

Table II. The comparison of different kernel size.

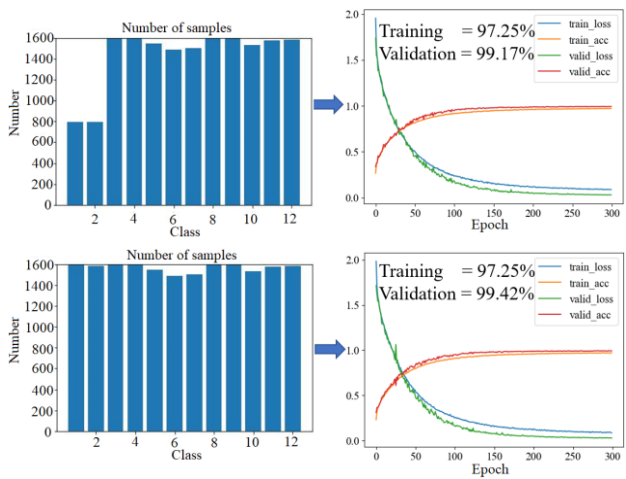

FIGURE 10. The comparison of different number of class 1 and 2.

Table III

| Dropout Rate | Accuracy(%) Training | Accuracy(%) Validation | Loss Training | Loss Validation |
|---|---|---|---|---|
| 0 | **100%** | 81.27% | **0.08** | 1.39 |
| 0.1 | **100%** | 86.70% | 0.10 | 0.71 |
| 0.2 | 99.72% | 88.35% | 0.14 | 0.53 |
| 0.3 | 99.40% | 87.34% | 0.17 | 0.49 |
| 0.4 | 98.84% | 88.19% | 0.19 | 0.44 |
| 0.5 | 97.87% | **89.89%** | 0.23 | 0.42 |
| 0.6 | 96.40% | 88.45% | 0.27 | **0.40** |
| 0.7 | 92.37% | 87.76% | 0.37 | 0.42 |
| 0.8 | 84.51% | 81.86% | 0.52 | 0.53 |

Table III. The comparison of different dropout rate.

### 2) KERNEL SIZE
In this experiment, an evaluation of the effect of different kernel sizes on performance was conducted. The evaluation was performed using raw data, and the attention network was not utilized. The results, as depicted in Table II, indicate that accuracy increases and loss decreases as the kernel size increases. A plateau in both accuracy and loss appears to be reached when the kernel size reaches 15. Based on these observations, kernel size 15 was selected for use in subsequent experiments.

### 3) DROPOUT
In this study, the impact of different dropout rates on performance during training in each TCN block was analyzed. The evaluation was performed on raw data, and the attention mechanism was not utilized. The results, as shown in Table III, indicate that the lowest loss was achieved when the dropout rate was set to 0.6, while the highest accuracy was attained with a dropout rate of 0.5. Based on these findings, the dropout rate was set to 0.5 during the training phase. It should be noted that the validation accuracy, as depicted in Table III, is expected to be higher than the training accuracy due to the operation

### 4) ATTENTION
In the examination of attention, the performance was evaluated at various layers while attention was applied. Three separate experiments were conducted to assess the effect of attention on the model. Experiment 1 involved applying attention before TCN, experiment 2 involved applying attention to the output of TCN and feeding it to FCN, and experiment 3 involved applying attention at the beginning of each TCN layer. As shown in Fig. 11, the performance was almost identical when attention was applied in experiments 1 and 2. However, experiment 3 was discarded due to its high computational demands, as high accuracy could be achieved with a single layer of attention. Ultimately, it was decided to apply attention based on experiment 1.

Furthermore, the performance with and without attention was compared. According to Fig. 12, the models with attention reached 99% accuracy around the 100th epoch, and this convergence was faster than the models without attention, which reached 99% accuracy around the 180th epoch. This result demonstrates the significance of attention in achieving fast convergence during WiFi signal processing.

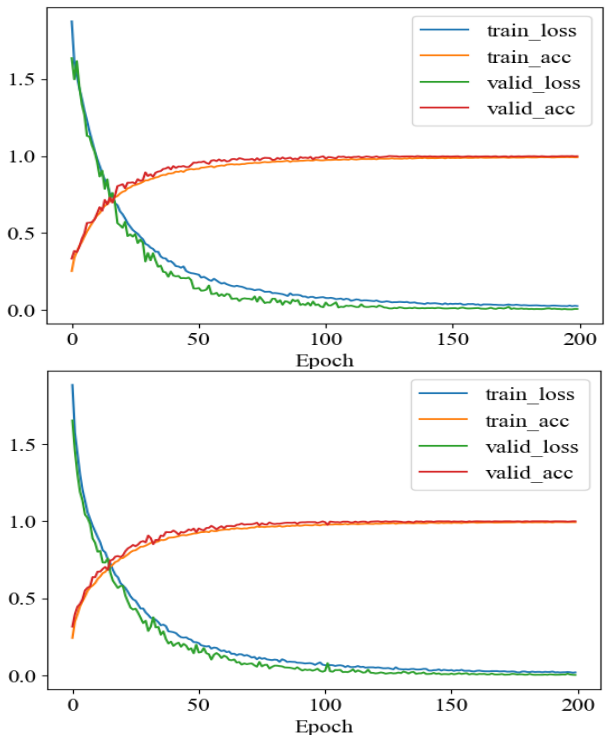

**FIGURE 11. The loss and accuracy curve of one-layer attention. The curve above is applying attention before TCN and the curve below is applying after TCN.**

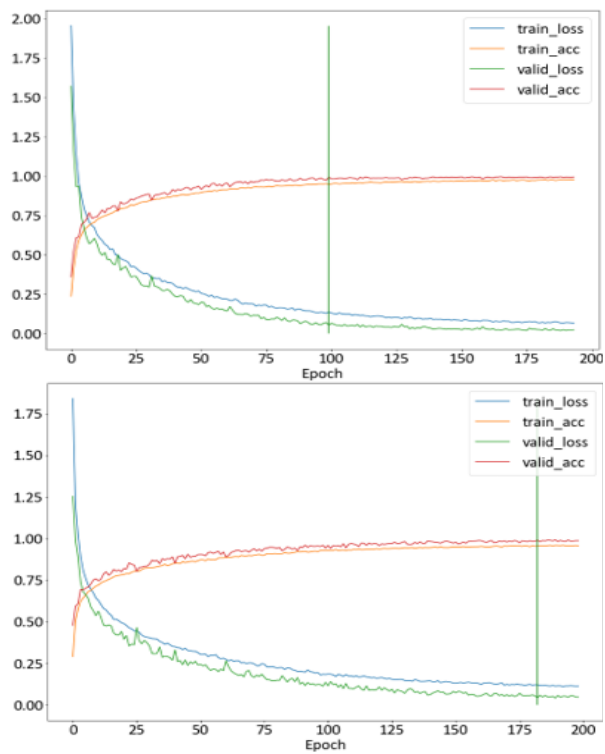

**FIGURE 12. The loss and accuracy curve with attention and without attention.**

## V. CONCLUSION AND FUTURE WORK

In this paper, a novel approach for recognizing human-human interactions using Wi-Fi signals is presented. The proposed method incorporates an augmentation method, an attention mechanism, and a temporal convolution network (TCN-AA) to efficiently extract features from time-sequence data while maintaining a low number of parameters. The experimental results demonstrate the superiority of the TCN-AA model over existing state-of-the-art methods, as it achieved a remarkable accuracy of 99.42% on the public dataset, outperforming the current best by 3%.

As future work, the applicability and adaptability of the proposed method in various environments, with different subjects, and for more complex behaviors will be investigated.

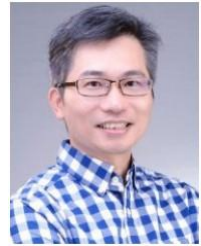

**Chih-Yang Lin** (Senior Member, IEEE) is presently affiliated with the Department of Mechanical Engineering at National Central University in Taoyuan, Taiwan, having previously served as a dean of International Academy, chief of Global affairs office, and a member of the Department of Electrical Engineering at Yuan-Ze University in Taoyuan, Taiwan. He has been recognized as an IET Fellow and has contributed to over 200 papers that have been featured in a wide range of international conferences and journals. His research interests are primarily focused on computer vision, machine learning, deep learning, image processing, big data analysis, and the design of surveillance systems.

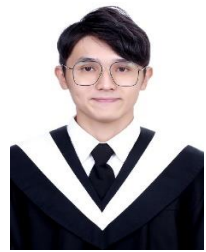

**Chia-Yu Lin** received a B.Sc. degree in Computer Science and Information Engineering from National Central University in Taiwan. He is pursuing a Master of Science degree in Computer Science and Information Engineering from National Central University in Taiwan. His research interests include human activity recognition, Wi-Fi-based recognition and few-shot learning.

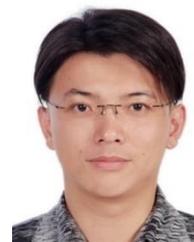

**Yu-Tso Liu** received the M.S. degree in Electrical Engineering from National Dong Hwa University, Taiwan. He is currently pursuing a Ph.D. degree in Computer Science and Information Engineering from National Central University, Taiwan. His research interests include wireless communication, the application of edge artificial intelligence, deep learning, and the Internet of Things. He is currently employed as a Senior Engineer and Product Testing Division Manager in Pantherun Technologies, where he specializes in the development and testing of industrial network communication products.

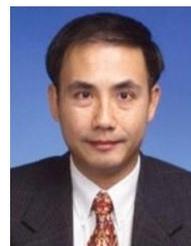

**Timothy K. Shih** (Senior Member, IEEE) is a Distinguished Professor at the National Central University (NCU), Taiwan. In NCU, he was the Vice Dean of College of EECS and the founding Director of Innovative AI Research Center. He was the Dean of the College of Computer Science, Asia University, Taiwan and the Chairman of the CSIE Department at Tamkang University, Taiwan. Prof. Shih is a Fellow of the Institution of Engineering and Technology (IET). He was also the founding Chairman Emeritus of the IET Taipei Local Network. In addition, he is a senior member of ACM and a senior member of IEEE. He was the founder and co-editor-in-chief of the International Journal of Distance Education Technologies, USA. He was the Associate Editor of IEEE Computing Now. And, he was the associate editors of the IEEE Transactions on Learning Technologies, the ACM Transactions on Internet Technology, and the IEEE Transactions on Multimedia. Prof. Shih was the Conference Co-Chair of the 2004 IEEE Internet. (http://tshih.minelab.tw)